\documentclass[runningheads]{llncs}
\usepackage{graphicx}
\begin{document}
\title{SlicerTMS: Real-Time Visualization of Transcranial Magnetic Stimulation for Mental Health Treatment}
\titlerunning{SlicerTMS}
\author{Loraine Franke\inst{1} \and Tae Young Park\inst{2,3} \and
Jie Luo\inst{2} \and Yogesh Rathi\inst{2} \and
Steve Pieper\inst{4} \and Lipeng Ning\inst{2}$^*$ \and Daniel Haehn\inst{1}$^*$}

\institute{University of Massachusetts Boston, Boston, USA, \email{\{franke,haehn\}@mpsych.org} \and
Harvard Medical School, Boston, USA, \email{\{typark,jluo5,yogesh,lning\}@bwh.harvard.edu} \and
Korea University of Science and Technology, Seoul, Republic of Korea \and
Isomics, Inc., Cambridge, USA, \email{pieper@isomics.com}\\
  \textsuperscript{*}Joint senior author
}
\authorrunning{Franke et al.}
\maketitle              
\begin{abstract}
We present a real-time visualization system for Transcranial Magnetic Stimulation (TMS), a non-invasive neuromodulation technique for treating various brain disorders and mental health diseases. Our solution targets the current challenges of slow and labor-intensive practices in treatment planning. Integrating Deep Learning (DL), our system rapidly predicts electric field (E-field) distributions in 0.2 seconds for precise and effective brain stimulation. The core advancement lies in our tool's real-time neuronavigation visualization capabilities, which support clinicians in making more informed decisions quickly and effectively. We assess our system's performance through three studies: First, a real-world use case scenario in a clinical setting, providing concrete feedback on applicability and usability in a practical environment. Second, a comparative analysis with another TMS tool focusing on computational efficiency across various hardware platforms. Lastly, we conducted an expert user study to measure usability and influence in optimizing TMS treatment planning. The system is openly available for community use and further development on GitHub: \url{https://github.com/lorifranke/SlicerTMS}.

\end{abstract}

\section{Introduction}
Transcranial Magnetic Stimulation (TMS)~\cite{barker1985non} is a non-invasive method to treat brain disorders such as depression, migraines, and addictions, and it is used in research for Parkinson's, Schizophrenia, and Alzheimer's disease. The technique uses a device called \emph{TMS coil} to stimulate brain neurons by generating electromagnetic fields (E-Field). Precise coil placement for targeted therapy is crucial for treatment outcomes~\cite{barbour2019individualized,bender2020customizing}. Despite its wide clinical application~\cite{nardone2012effect,carpenter2012transcranial,antonelli2021transcranial}, traditional E-field estimation methods involve complex head models and computationally demanding techniques, which are not practical for real-time clinical use due to their dependence on high-performance graphics processing units (GPUs)~\cite{sparing2008transcranial,gomez2021fast}. This complexity leads to lengthy preparation times, including many hours of manual setup and automated model construction~\cite{weise2023precise}. The need for real-time visualization for TMS was already identified in 2004~\cite{noirhomme2004registration} but poses challenges for real-time prediction and visualization in clinical settings. 

In this paper, we present SlicerTMS, the first open-source software enabling real-time E-field prediction and visualization for TMS treatment, using deep learning (DL) within a neuronavigation system for immediate visualization. This innovation allows clinicians to adjust the TMS coil on a patient's head in real-time, with the DL model providing instant E-field updates, a significant advancement over traditional TMS tools that offer only static visualizations~\cite{souza2018development,afuwape2020measurement,saturnino2019simnibs}. In clinical settings, the demand for real-time visualization and AI for enhanced clinical planning and medical imaging analysis is growing, yet there's a gap between AI advancements and their application in healthcare. Besides real-time visualizations, traditional TMS tools often lack deep learning and integration with neuronavigation~\cite{leuze2018mixed,schutz2022audiovisual,sathyanarayana2020comparison}. Furthermore, SlicerTMS utilizes web-based AR for enhanced interaction and placement of TMS coils and MRI images, benefiting from the advancement of XR technologies in medical applications~\cite{stramba2020transcranial}. We enhance the system's usability and accuracy by incorporating 3D Slicer~\cite{fedorov20123d}, a widely used open-source visualization platform for medical imaging and optical tracking. Our integration accelerates E-field estimation and leverages the robust capabilities of 3D Slicer, which various tools have extended in recent years~\cite{park2022application,preiswerk2019open,siddiqui2023interactive,pinter2020slicervr}. We evaluate SlicerTMS's usability and functionality through experiments on various hardware and data from ten patients, comparing it with a current TMS tool to demonstrate speed and usability advantages. A real-world use case in a TMS clinic shows the system's practicality, complemented by feedback from a domain expert study. Involving medical experts for user-centered design is critical to bridge this gap and effectively integrate and optimize AI technologies in clinics as real-world applications. SlicerTMS enhances TMS visualization and medical visualization by providing a novel, open-source platform for real-time E-field visualization and deep learning, facilitating efficient and precise clinical brain stimulation planning. 

\section{Implementation}

\subsection{Neuronavigation Visualization Component}

\begin{figure}[t]
    \centering
    \includegraphics[width=0.85\linewidth]{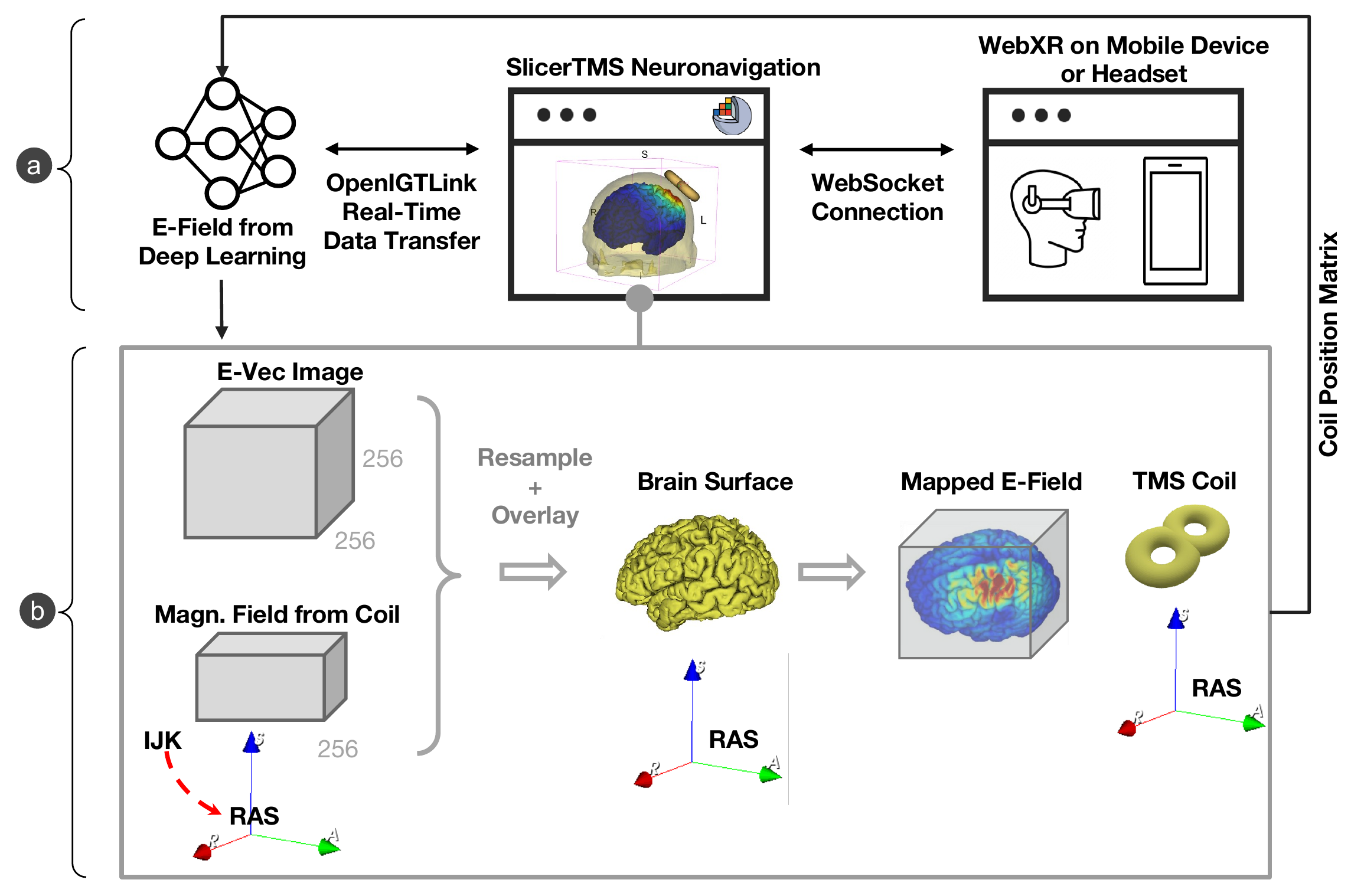}
    \caption{\textbf{a) Components}: Neural Network (left) predicts E-field and transfers it to SlicerTMS via OpenIGTLinkIF. WebSocket supports browser connection to WebXR to interact with visualization in AR (right). \textbf{b) Neuronavigation Visualization:} Incoming magnetic vector field images are transformed according to coil position, then overlayed with the brain mesh. Consistent rotation of vector direction in each voxel as rigid transform is critical. The 3D coil can be moved interactively while sending new coil positioning matrices back to neural network generating a new field.}
    \label{fig:workflow}
\end{figure}
\begin{figure}
    \centering
    \includegraphics[height=3cm]{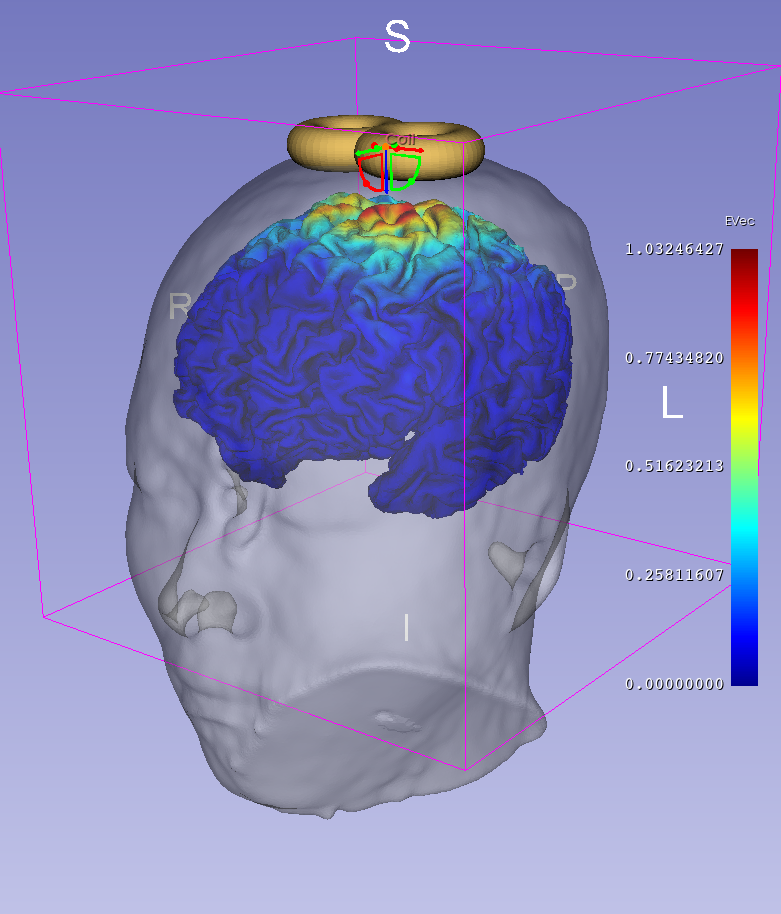}
    \includegraphics[height=3cm]{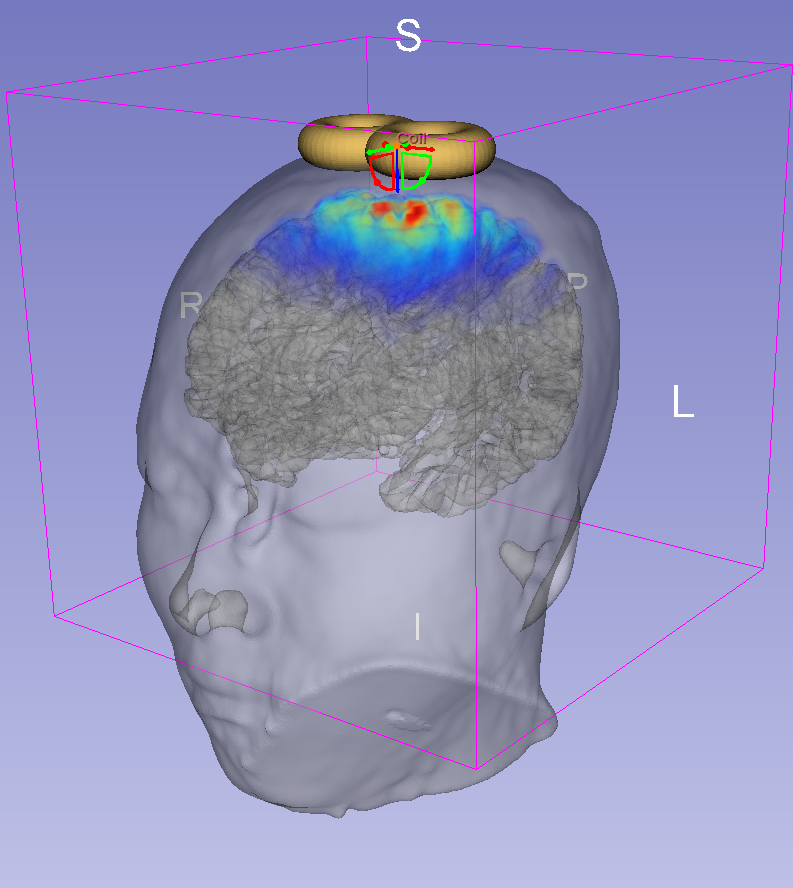}
    \includegraphics[height=3cm]{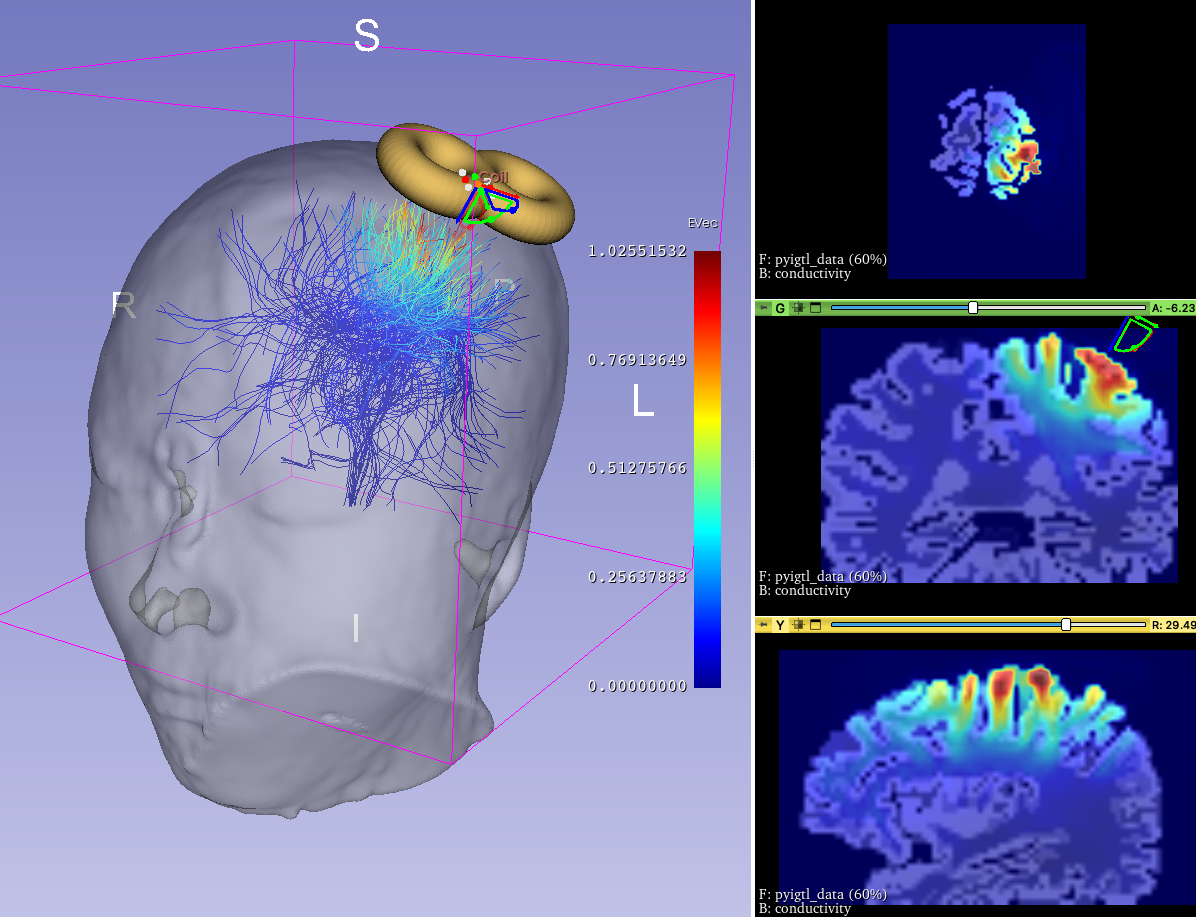}
    \caption{\textbf{Gallery of example visualizations: Brain Mesh vs. Volume Rendering vs. Fiber Tractography in SlicerTMS.} E-field in 3D on gray matter with a figure-8 coil (\textbf{Left}), E-field on MRI volumetric data (\textbf{Center}), and E-field on full-brain tractography fibers with adjustable ROI and 2D slices in various directions (\textbf{Right}).}
    \label{fig:renderings}
\end{figure}

\vspace{-.3cm}
The final interface of SlicerTMS (Figure \ref{fig:workflow}) integrates a client-server architecture with neural networks for electric field predictions, a primary user interface, and augmented reality visualization.
SlicerTMS provides real-time rendering of predicted E-Field on different modalities: a 3D brain mesh, its 3D volume, and the brain's white matter fiber bundles obtained from MRI scans for patient-specific modeling. Our tool is integrated into the neuronavigation software 3D Slicer~\cite{fedorov20123d}, with Kitware's~\href{https://vtk.org/}{VTK} framework to manipulate graphical elements. Users can visualize and explore real-time TMS results from standard desktop monitors on different operating systems for easy navigation with adjustable visualization parameters for a thorough analysis of different scenarios. Data is transmitted from the neural network through a protocol called \textit{OpenIGTLink}, a tool for network communication with external software or hardware using a protocol allowing real-time image and position streaming with submillisecond latency up to 1024 fps~\cite{tokuda2009openigtlink}, for performance and scalability criteria. The neural network server component can run locally or as a remote service on a GPU server with data transfer via OpenIGTLink and facilitate smooth performance during computationally intensive tasks. Each time the neural network generates a new E-Field, it updates the TMS module and visualization. Inside SlicerTMS' user interface, users can manually adjust the coil by dragging and rotating it with the cursor into the desired position or using text input fields to enter a coil position matrix where the 3D coil automatically jumps into position. While moving the coil, we inform the neural network via OpenIGTLink, generating a new E-Field for this specific coil position (see Figure~\ref{fig:workflow}). We created a coil as a 3D mesh in .stl format, simulating a 'figure 8' coil like the Magstim-70mm-Figure8, a standard TMS treatment coil. Users can freely use any other type of TMS coil by exchanging the example coil with their coil file in the data folder. Equally, all other files, such as skin, gray matter, and conductivity files, can be exchanged individually, allowing for patient-specific treatment. 2D and 3D views of the neuronavigation component are illustrated in Figure~\ref{fig:renderings}.

\vspace{-0.2cm}

\subsection{Deep Learning Pipeline for E-Field Prediction}
Several deep-learning-based approaches have been developed for estimating brain E-Fields~\cite{yokota2019real,xu2021rapid,li2022computation,stenroos2019real,daneshzand2021rapid}. While ~\cite{yokota2019real} did not consider subject-specific brain connectivity, methods of~\cite{xu2021rapid} and ~\cite{li2022computation} use subject-specific whole-brain E-field prediction. We train a multi-scale 3D-Res-UNet model with a reduced field-of-view to accelerate prediction time. We use $\mathrm{T}_{1w}$ MRI images and diffusion MRI from the Human Connectome Project (HCP)~\cite{van2012human} for training. Using SimNIBS~\cite{saturnino2019simnibs}, we constructed volume conductor models from $\mathrm{T}_{1w}$ images and estimated anisotropic tissue conductivity tensors from diffusion MRI of $b=0$ and $1000 s/mm^2$ volumes. 
We varied the coil's position and orientation for diverse E-field maps, aligning with EEG 10-10 system positions and adjusting the coil handle to 78 different directions. We randomly selected a total of 300 cases from this dataset. Models were trained on NVIDIA V100 32GB GPU with PyTorch for 2000 epochs until convergence with Adam optimizer and a tailored learning rate strategy. Input volumes for the models comprised concatenated data of conductivity tensors and time derivative of the magnetic potential, formatted as three-dimensional arrays to preserve spatial information crucial for E-field prediction. Performance evaluations were based on normalized root-mean-square error (NE). Our model has comparable accuracy reported by~\cite{xu2021rapid}, achieving a NE of $0.198 \pm 0.017$. 

\subsection{AR Component with WebXR}

SlicerTMS features a web server for connection to local browsers, using \href{https://www.tornadoweb.org/}{Tornado} for secure WebSocket communication to enable real-time interaction with the 3D Slicer neuronavigation platform. Users can control the coil by moving or rotating their mobile phone, which has a depth sensor, with position updates sent via WebSocket. The system uses JavaScript with \href{https://threejs.org/}{ThreeJS} for client-side rendering and the WebXR API~\cite{franke2020modern} supporting mobile-based AR and VR headsets, offering three coil interaction options (see Figure 1 Supplemental Material).
\vspace{-.1cm}

\section{Real-world Use Case Scenario}\label{sec:study1} 
For further enhancements, we tested our prototype in a Boston area clinic. This out-of-lab study involved six participants, including clinicians, a subject undergoing an MRI scan, and researchers, to simulate TMS treatment using an optical tracker system for coil placement. Collaborating with experts, we faced challenges aligning brain images with the patient's head using the optical tracker. We must ensure a reliable internet connection to run our deep learning model. This hands-on testing provided valuable real-world insights into the usability and integration of SlicerTMS in clinical settings. We included these improvements into our prototype for efficient TMS clinic workflow, including UI improvements for manual coil position entry, brain fibers visualization from diffusion MRI data for clinical and patient-specific treatment relevance~\cite{norton2017slicerdmri,franke2021fiberstars}, color map updates for electric field distributions, mobile device-enabled AR feature for setup simplification, and electric field visualization on volume data to aid radiological studies, as in Figure~\ref{fig:renderings}. The updates aim to refine SlicerTMS for smoother, more effective TMS treatment planning.
\begin{figure}[h]
    \centering
    \includegraphics[width=0.8\linewidth]{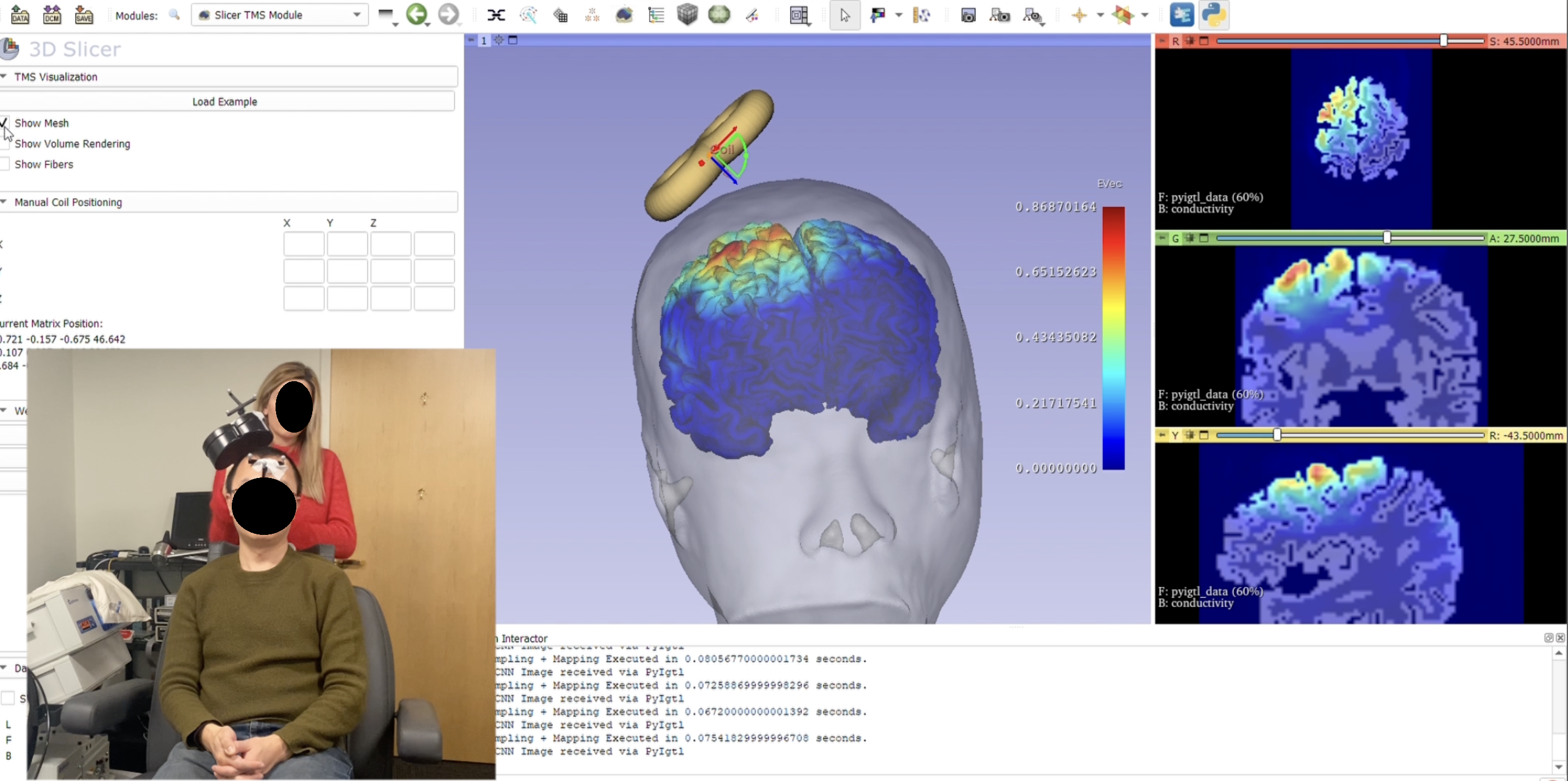}
    \caption{\textbf{TMS Clinic and Output in SlicerTMS.} Researcher administering treatment by standing behind patient and adjusting the TMS coil to target the right brain areas, meanwhile SlicerTMS is running in with a remote connection to server to predict the electric field in real-time on the brain mesh shown in the UI.}
    \label{fig:clinic}
\end{figure}
\vspace{-0.1cm}

\section{Performance Evaluation}

\subsection{Technical Performance Experiments of SlicerTMS}
For experiments, we randomly selected ten different subjects from the Human Connectome Project (\textit{HCP}) dataset~\cite{van2012human} consisting of so-called T1$_w$ MRIs. T1$_w$ MPRAGE images were acquired with 0.7mm isotropic voxels. We replaced each subject's conductivity file, skin, and brain mesh while the coil and pre-trained deep learning model remained the same throughout all subjects.  We evaluated SlicerTMS on four devices for real-time visualization with deep learning and compared it to a leading TMS visualization tool in Section~\ref{sec:comparison}.
We evaluated based on E-field prediction (CNN) and real-time visualization (Vis). Results can be seen in Table~\ref{tab:performance}. Input data consisted of an electric field of shape 70x90x50x1 generated by our trained model. The visualization includes resampling Nifti images and simultaneously projecting the E-field on a 3D brain mesh, brain volume, and tractography data. We registered how long the code needs to execute each \textit{run}, where a run is defined as the coil movement triggering the neural network, resulting in predicting an E-field that is immediately visualized in SlicerTMS. To get a precise measurement, we averaged the time of 50 runs on following devices: an Apple M1 MacBookBook Air with 16GB Memory (Apple M1), a workstation computer with an Intel Core i9-9980XE with 36 CPUs @ 3.00GHz and 64 GB RAM (CPU i9), an NVIDIA GeForce RTX 2080 GPU (2080Ti), and a remote NVIDIA A100 GPU (A100). Results of our performance testing indicate that our neural network runs on average in less than $0.2$ seconds and real-time visualization in less than ten milliseconds. 

\begin{table}[h!]
\footnotesize 
\setlength{\tabcolsep}{1.5pt} 
\vspace{-.1cm}
    \centering
    \caption{\textbf{Performance Evaluation.} Timings for E-field prediction (CNN) and visualization (Vis.) are shown separately for ten different subjects using four different hardware configurations. The remote A100 setup using cloud-based inference is fastest.
    All times in seconds.}
    \label{tab:performance}
    \resizebox{\linewidth}{!}{
    \begin{tabular}{ll|c|c|c|c|r}
\toprule
        && \textbf{Apple M1} & \textbf{CPU i9 } & \textbf{2080Ti} & \textbf{A100 (Remote)} &{\textbf{Mean[s]}} \\
        & & Mean[s] $\pm$std. & Mean[s]$\pm$std. & Mean[s]$\pm$std. & Mean[s]$\pm$std. & ~Mean[s]$\pm$std. \\
        \midrule
\textbf{Subject 1}&
\textbf{CNN}& $3.03894\pm0.11715$ & $0.16642\pm0.16238$& $\textbf{0.04028}\pm0.00565$ & $0.05592\pm0.01595$ & $0.82539\pm0.00773$ \\
&\textbf{Vis.}& $\textbf{0.04755}\pm0.0019$ & $0.09336\pm0.01284$ & $0.09555\pm0.00825$ & $0.09428\pm0.01050$ & $0.08254\pm0.07528$\\
\midrule
\textbf{Subject 2}&
\textbf{CNN}& $3.04025\pm0.07488$ & $0.17434\pm0.01121$ & $\textbf{0.06735}\pm0.18290$ & $0.06934\pm0.01456$ & $0.83398\pm0.07089$ \\
&\textbf{Vis.}& $\textbf{0.04682}\pm0.00084$ & $0.09355\pm0.01257$ & $0.09865\pm0.00683$ & $0.09886\pm0.01168$ & $0.08448\pm0.00798$\\
\midrule
\textbf{Subject 3}&
\textbf{CNN}& $3.03730\pm0.0521$ & $0.16703\pm0.00911$ & $\textbf{0.06516}\pm0.17955$ & $0.06935\pm0.09588$ & $0.83471\pm0.08416$  \\
&\textbf{Vis.}& $0.04640\pm0.00283$ & $0.0949\pm0.01431$ & $0.09591\pm0.01053$ & $0.09699\pm0.01376$ & $0.08355\pm0.01036$ \\
\midrule
\textbf{Subject 4}&
\textbf{CNN}& $3.00502\pm0.04971$ & $0.16979\pm0.00839$ & $0.06585\pm0.17467$ &  $\textbf{0.05568}\pm0.01793$ & $0.82409\pm0.06268$  \\
&\textbf{Vis.}& $\textbf{0.05050}\pm0.01596$ & $0.09094\pm0.00856$ & $0.09528\pm0.01075$ & $0.09636\pm0.01216$ & $0.08327\pm0.01186$ \\
\midrule
\textbf{Subject 5}&
\textbf{CNN}& 3.04525$\pm$0.06020 & 0.17424$\pm$0.00842 & \textbf{0.0674}$\pm$0.18280 & 0.06773$\pm$0.09635 & 0.83866$\pm$0.08694 \\
&\textbf{Vis.}& \textbf{0.04997}$\pm$0.01436 & 0.09611$\pm$0.01756 & 0.09684$\pm$0.00724 & 0.09721$\pm$0.01347 & 0.08503$\pm$0.01316 \\
\midrule
\textbf{Subject 6}&
\textbf{CNN}& 3.02331$\pm$0.06009 & 0.17044$\pm$0.01210 & 0.06975$\pm$0.19970 & \textbf{0.06397}$\pm$0.08236 & 0.83187$\pm$0.08856 \\
&\textbf{Vis.}& \textbf{0.04699}$\pm$0.00096 & 0.09695$\pm$0.02807 & 0.09727$\pm$0.01152 & 0.09359$\pm$0.01363 & 0.08370$\pm$0.01355 \\
\midrule
\textbf{Subject 7}&
\textbf{CNN}& 3.00740$\pm$0.05809 & 0.17397$\pm$0.00812 & \textbf{0.06340}$\pm$0.17313 & 0.06527$\pm$0.07967 & 0.82752$\pm$0.07975  \\
&\textbf{Vis.}& \textbf{0.04950}$\pm$0.01188 & 0.09514$\pm$0.01094 & 0.09644$\pm$0.01840 & 0.09620$\pm$0.01194 & 0.08432$\pm$0.01329 \\
\midrule
\textbf{Subject 8}&
\textbf{CNN}& 3.03198$\pm$0.04614 & 0.16830$\pm$0.01014 & 0.06624$\pm$0.17561 & \textbf{0.05417}$\pm$0.01574 & 0.83017$\pm$0.06191  \\
&\textbf{Vis.}& \textbf{0.04752}$\pm$0.00297 & 0.09478$\pm$0.01425 & 0.09733$\pm$0.00740 & 0.09691$\pm$0.01182 & 0.08413$\pm$0.00911 \\
\midrule
\textbf{Subject 9}&
\textbf{CNN}& 3.0257$\pm$0.05572 & 0.17037$\pm$0.00988 & 0.0657$\pm$0.17642 & \textbf{0.06445}$\pm$0.07943 & 0.83156$\pm$0.08037  \\
&\textbf{Vis.}& \textbf{0.04698}$\pm$0.0008 & 0.09527$\pm$0.01315 & 0.09443$\pm$0.00967 & 0.09537$\pm$0.01228 & 0.08302$\pm$0.00899 \\
\midrule
\textbf{Subject 10}&
\textbf{CNN}& 3.0236$\pm$0.05616 & 0.16972$\pm$0.01004 & \textbf{0.06485}$\pm$0.17405 & 0.06586$\pm$0.07994 & 0.83101$\pm$0.08005  \\
&\textbf{Vis.}& \textbf{0.05375}$\pm$0.02470 & 0.09406$\pm$0.01384 & 0.09503$\pm$0.01074 & 0.09631$\pm$0.01201 & 0.08479$\pm$0.01534 \\
\midrule
\textbf{Mean [s]}& \textbf{both} & 3.08655$\pm$0.07076 & 0.26498$\pm$0.03959 &  0.15987$\pm$0.01553 & \textbf{0.15784}$\pm$0.07011 & 0.91479$\pm$0.08826\\
    \bottomrule
    \end{tabular}}
\end{table}
\vspace{-0.2cm}

\subsection{Comparative Performance Analysis}\label{sec:comparison}

\begin{figure}
    \centering
    \includegraphics[width=0.3\linewidth]{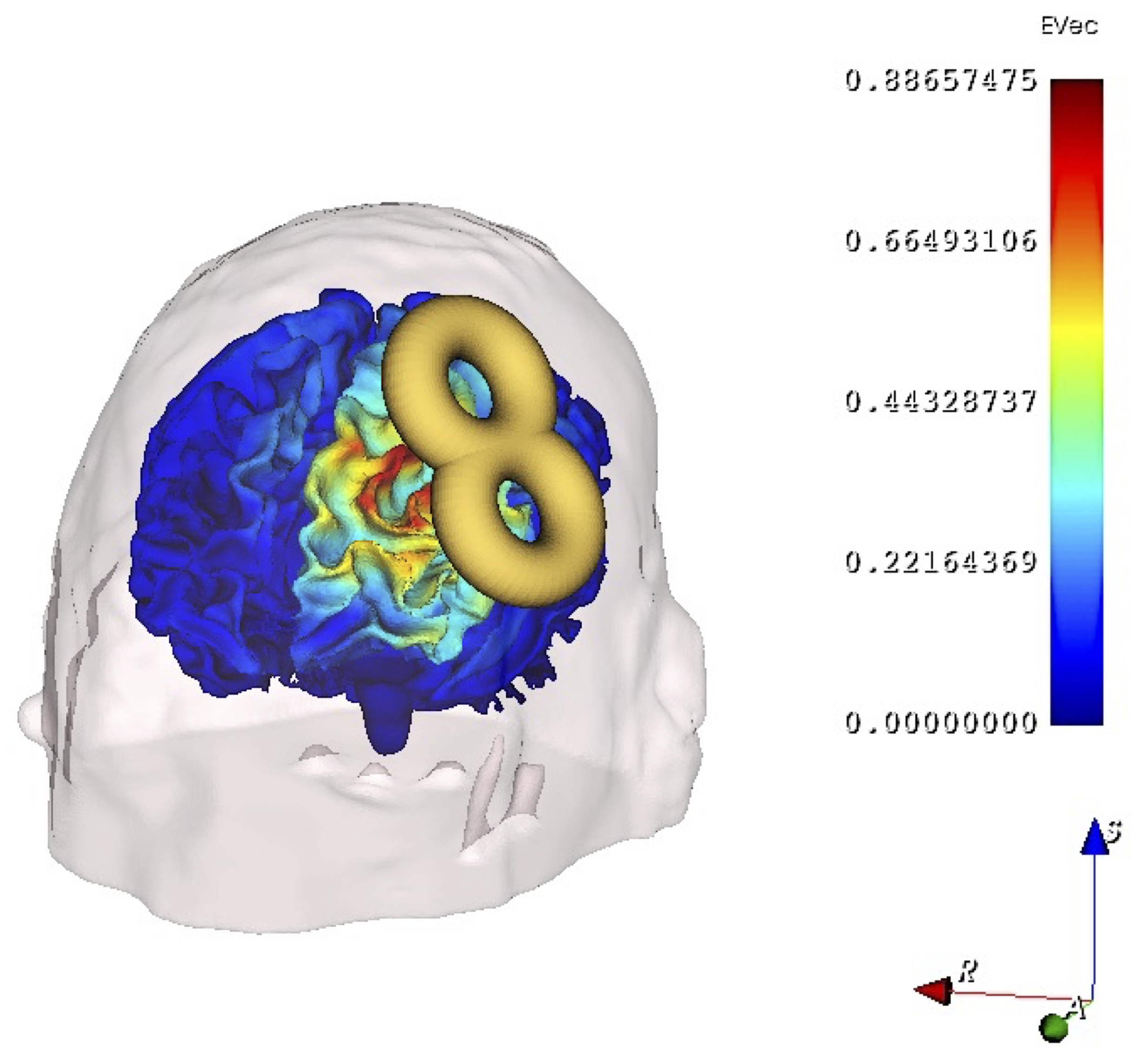}
    \hspace{1.8cm}
    \includegraphics[width=0.3\linewidth]{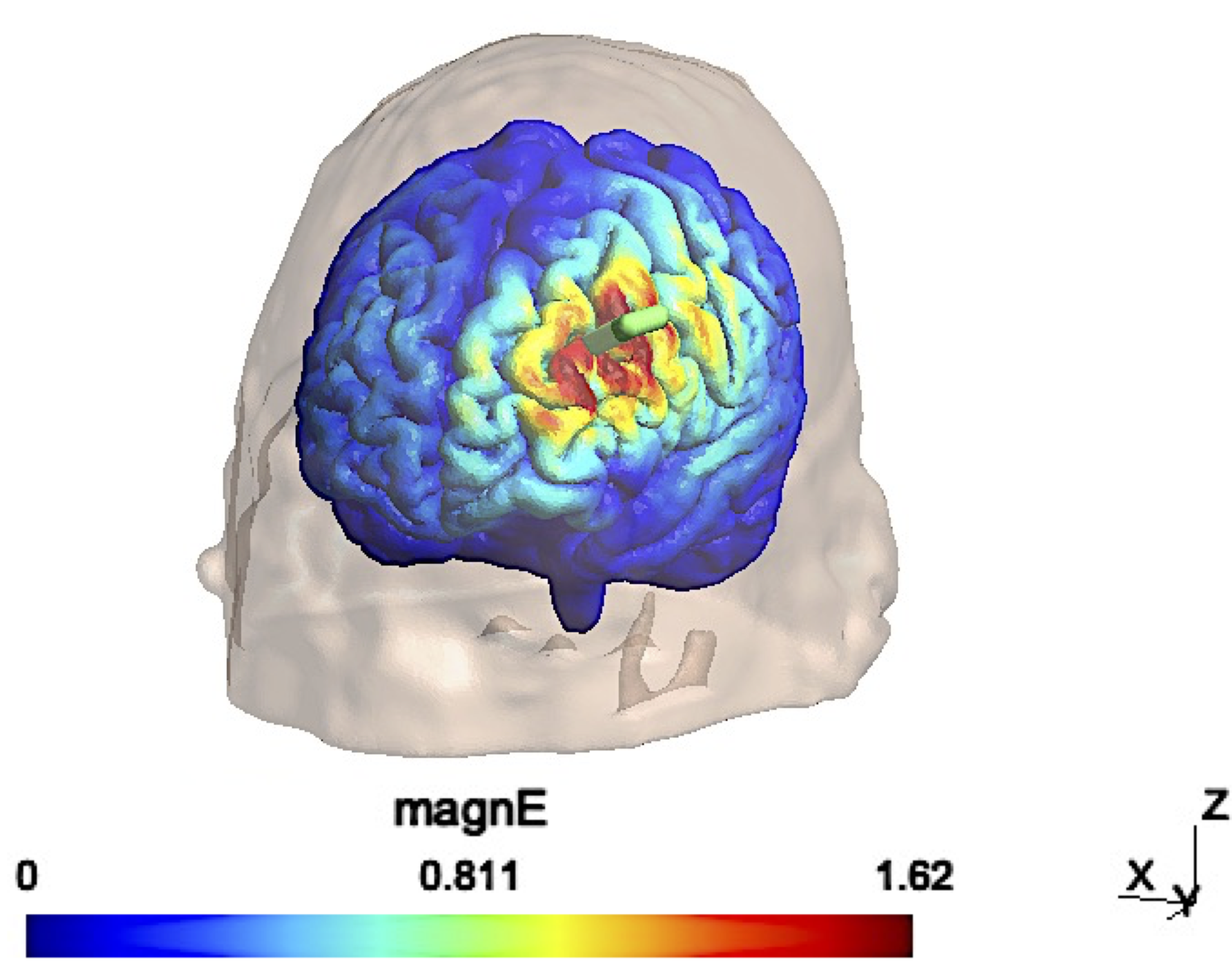}
    \caption{\textbf{SlicerTMS vs. SimNIBS E-Field Visualization.} \textbf{Left:} SlicerTMS E-field with movable Figure 8 coil, color legend indicates strength of  E-field and axes with directions. \textbf{Right:} SimNIBS E-field on brain surface with coil direction (green handle).}
    \label{fig:simnibsvsslicer}
\end{figure}

We compared SlicerTMS to the existing TMS visualization software  SimNIBS\cite{saturnino2019simnibs}. 
Unlike SlicerTMS, SimNIBS does not rely on deep learning or real-time but solely on statically visualizing E-fields based on manual coil placement. We tested both tools on ten subjects from the HCP dataset. We used SimNIBS v4.0 to compute E-fields induced by a Magstim-70mm-Figure8 coil at various locations and orientations. We measured the time only for the visualization part inside SimNIBS. We created the E-field for each of the ten subjects in random positions and entered these exact coil position matrices in SlicerTMS to generate the same E-fields. Figure~\ref{fig:simnibsvsslicer} shows the same coil positions in both tools. We did not measure additional times SimNIBS requires, i.e., calculating the dA/dt field, computing matrices, or solving the system. Table~\ref{tab:comparison} shows the results of visualization speed comparison on two CPU machines. SlicerTMS took, on average, $0.08506$ seconds, while SimNIBS needs $7.58798$ seconds to visualize an E-field on the brain. We conducted a two-sided t-test to determine significant differences in group means. The null hypothesis states no significant difference between SlicerTMS and SimNIBS (\textbf{H$_0$}), while the alternative hypothesis states a significant difference between groups (\textbf{H$_1$}). We used an independent samples t-test as the two groups are independent. The t-statistic is $t_{38} = 56.3$, $p<0.0001$, indicating a significant difference between groups. We reject \textbf{H$_0$} and accept the alternative hypothesis that there is a significant difference between SlicerTMS and SimNIBS.

\begin{table}[h!]
\footnotesize 
\setlength{\tabcolsep}{1pt} 
\begin{center}
\caption{\textbf{Comparison with SimNIBS.} We measure visualization speed of an E-field on the brain mesh at fixed TMS coil positions in both tools. We report measurements for two hardware configurations. All times are in seconds. SlicerTMS is over 78$\times$ faster.}
    \label{tab:comparison}
    \resizebox{\linewidth}{!}{
    \begin{tabular}{l|ccc|ccc|r}
& \multicolumn{3}{c}{\textbf{SimNIBS}} & \multicolumn{3}{c}{\textbf{SlicerTMS}} & \\
& \textbf{Apple M1 [s]} & \textbf{CPU i9 [s]} & {\textbf{~Mean [s]}}& \textbf{Apple M1 [s]} & \textbf{CPU i9 [s]} & {\textbf{~Mean [s]}}  & {\textbf{Improvement}} \\
    \midrule
    \textbf{Subject 1}& 6.81622 & 6.43095 & 6.62369 & 0.05539 & 0.09607 & 0.09607& 66.67x faster \\
    \textbf{Subject 2}& 8.22622 & 7.98845 & 8.10733 & 0.06353 & 0.09891 & 0.09891& 81.97x faster \\
    \textbf{Subject 3}& 7.69764 & 7.37311 & 7.53538 & 0.05733 & 0.09138 & 0.09138 & 82.46x faster\\
    \textbf{Subject 4}& 7.08191 & 7.08276 & 7.08233 & 0.08974 & 0.10517 & 0.10517 & 67.34x faster\\
    \textbf{Subject 5}& 7.72296 & 6.60987 & 7.16642 & 0.12807 & 0.09652 & 0.09653& 74.24x faster\\
    \textbf{Subject 6}& 7.91043 & 7.43637 & 7.67339 & 0.04735 & 0.09448 & 0.09448 & 81.21x faster\\
    \textbf{Subject 7}& 8.80702 & 8.10869 & 8.45786 & 0.06146 & 0.10163 & 0.10163 & 83.22x faster\\
    \textbf{Subject 8}& 7.90561 & 7.3814 & 7.64351 & 0.05229 & 0.09645 & 0.09645 & 79.25x faster\\
    \textbf{Subject 9}& 8.1296 & 8.0317 & 8.08065 & 0.12303 & 0.09769 & 0.09769 & 82.71x faster\\
    \textbf{Subject 10}&  7.2251 & 7.79360 & 7.50936 & 0.06049 & 0.084202 & 0.08420 & 89.18x faster\\
    \midrule
    \textbf{Mean[s]} & 7.75227 & 7.42369 & & 0.07387 & 0.09625&& \\
    \midrule
    \textbf{Mean[s]$\pm$std.}& \multicolumn{2}{c}{7.58798 $\pm$ 0.59553 } & & \multicolumn{2}{c}{0.08506 $\pm$ 0.02365} && \textbf{78.83x faster}\\
    \bottomrule
    \end{tabular}}
\end{center}
\vspace{-0.3cm}
\end{table}

\vspace{-0.1cm}

\section{Expert User Study}\label{sec:study2}
We conducted an expert user study to evaluate the impact of real-time TMS E-Field visualization on usability, workflow efficiency, and clinician trust in DL predictions, involving four TMS experts in tasks that simulate real-world scenarios. Feedback and a post-study questionnaire, alongside a NASA-TLX survey, were used to refine SlicerTMS, assessing its usability, utility, and workload in clinical settings. The study questionnaire can be found in the supplemental material.
The study results highlight the effectiveness of SlicerTMS in facilitating the interpretation of TMS electric fields, with experts successfully identifying brain regions and electric field strengths. Feedback indicates that SlicerTMS was appreciated for its functionality and ease of use, with areas identified for enhancement in 2D slice interactions. Participants noted SlicerTMS could improve TMS treatment planning and research by enabling real-time electric field visualization, a clear improvement over previous static images, and voiced interest in its use in clinical settings, along with suggestions for further improvements.

\section{Discussion}\label{sec:discussion}
SlicerTMS enhances brain stimulation treatment planning with real-time visual feedback. It is faster than previous methods, achieving near-smooth frame rates with remote GPUs. The system's dynamic visualizations improve the intuitive understanding of electric fields. TMS experts believe it could speed up treatment planning but suggest improving coil manipulation and 2D interactions. While acknowledging SlicerTMS's advancements, we also recognize the limitations compared to tools like SimNIBS, aiming to incorporate additional functionalities to bridge these gaps in future developments.

\section{Conclusion and Future Work}
This paper presents SlicerTMS, an innovative AI system for real-time Transcranial Magnetic Stimulation visualization. It enhances coil placement through integration with 3D Slicer and augmented reality. Its effectiveness, validated in a TMS clinic and through professional feedback, highlights its potential to upgrade treatment planning with high-speed, flexible visualizations. Leveraging both local and remote GPUs, SlicerTMS advances medical visualization significantly. Future improvements will measure coil-cortex distance, electric field direction visualization, and enhanced augmented reality brain projections.


%
%
\bibliographystyle{splncs04}
\bibliography{bib}
\end{document}